\documentclass[a4paper]{article}

\begin {document}

\title {\bf Reply to "Comments on Bouda and Djama's 'Quantum Newton's 
Law' " }
\author{A.~Bouda\footnote{Electronic address: 
{\tt bouda\_a@yahoo.fr}} \ and T.~Djama\footnote{Electronic address: 
{\tt djama.toufik@caramail.com}}\\
Laboratoire de Physique Th\'eorique, Universit\'e de B\'eja\"\i a,\\ 
Route Targa Ouazemour, 06000 B\'eja\"\i a, Algeria\\}

\date{\today}

\maketitle

\begin{abstract}
\noindent
In this reply, we hope to bring clarifications about  the 
reservations expressed by Floyd in his comments, give 
further explanations about the choice of the approach and 
show that our fundamental result can be reproduced by other 
ways. We also establish that Floyd's trajectories manifest 
some ambiguities related to the mathematical choice of the couple 
of solutions of Schr\"odinger's equation. 
\end{abstract}

\vskip\baselineskip



\vskip\baselineskip

\newpage

In Floyd's comments \cite{Floyd1} on our previous paper \cite{BD1}, 
after having showed that  
\begin {equation}
2Et=S_0\; ,
\end {equation} 
it is stated that our equation of motion is the quantum reduced 
action. Firstly, we indicate that the above relation, up to an 
additive constant, is already written in our paper \cite{BD1}, 
Eq. (39). Secondly, relation (1) is valid only in the particular 
free particle case and it does not work for other 
potentials. Therefore, we can not assert that our equation of 
motion is the quantum reduced action. With regard to our velocity, 
it is an instantaneous velocity of the particle which is localized 
in space at each time. Furthermore, the knowledge of all the 
integration constants, even the non-classical ones $a$ and $b$, 
determines univocally the trajectory and the velocity at each 
time. 

Concerning the representation of the Hamilton's principal 
function $S$ as an integral of a Lagrangian, 
Floyd claimed that his finding can be generalized for 
the case where $\hbar$ is not considered close to 0. From 
the quantum Hamilton principal function, he proposed the 
Lagrangian 
\begin {eqnarray}
L(x,\dot{x},\ddot{x},\dot{\ddot{x}})={\partial S_0 
\over \partial x}\dot{x}-
{1\over 2m} \left({\partial S_0 \over \partial x}\right)^2 - 
V(x) \hskip39mm&& \nonumber\\
+{\hbar^2\over 4m}  \left[{3\over 2}\left( 
{\partial S_0 \over\partial x}\right)
^{- 2 }\left({\partial^2 S_0 \over \partial x^2}\right)^2-
\left( {\partial S_0 \over \partial x  }\right)^{- 1 }
\left({\partial^3 S_0 \over \partial x^3  }\right) \right]\; ,
\end {eqnarray}
and suggested that one should start from some of the relations in 
Ref. \cite{FM1} giving the derivatives of $S_0$ with respect to 
$x$ in terms of temporal derivatives of $x$. He added that the 
resulting $L(x,\dot{x},\ddot{x},\dot{\ddot{x}})$ and the resulting 
Lagrange equation are cumbersome. We would like to indicate 
that it is not only cumbersome to express the Lagrangian 
but Faraggi-Matone's relations, as given in \cite{FM1}, 
do not allow us to express the Lagrangian 
(2) as a function of $(x,\dot{x},\ddot{x},\dot{\ddot{x}})$. 
In fact, these relations are
\begin {eqnarray}
P & = & m \left( 1-{\partial Q \over \partial E} \right)
        {\dot x} \ , \nonumber\\
{\partial P \over \partial x} & = &
              -m{\partial^2 Q \over \partial x \partial E}{\dot x}
      + m \left( 1-{\partial Q \over \partial E} \right)
       {{\ddot x} \over {\dot x}} \ , \nonumber\\
{\partial^2 P \over \partial x^2} & = & 
     - m {\partial^3 Q \over \partial x^2 \partial E}{\dot x}
     -2m {\partial^2 Q \over \partial x \partial E}
     {{\ddot x} \over {\dot x}}
     + m \left( 1-{\partial Q \over \partial E} \right)
    \left( {{\dot{\ddot x}} \over {\dot x}^2}
     -{{\ddot x}^2 \over {\dot x}^3} \right) \ , \nonumber
\end {eqnarray}
in which the quantum potential, $Q$, must be substituted by
\begin {equation}
Q={\hbar^2 \over 4m}
  \left[ 
       {1 \over P}{\partial^2 P \over\partial x^2}
       - {3 \over2 }\left({1 \over P}
         {\partial P \over \partial x}\right)^2
  \right]
\end {equation}
and $P=\partial S_0 / \partial x$ is the conjugate momentum. 
It is clear that $\dot{x}$, $\ddot{x}$ and $\dot{\ddot{x}}$ 
are related to $P$, $\partial P / \partial x$, ..., 
$\partial^4 P / \partial x^4$ and $\partial P / \partial E$, 
..., $\partial^5 P / \partial x^4 \partial E$. In our point 
of view, it is not possible to express 
$(P, \partial P / \partial x, \partial^2 P / \partial x^2)$,  
and therefore the Lagrangian (2), only in terms of $(x, \dot{x}, 
\ddot{x}, \dot{\ddot{x}})$. In addition, the constant $E$ 
will appear in (2) and will be redundant. In order to avoid 
the above higher derivatives, an alternative is to use the 
solution of the quantum stationary Hamilton-Jacobi equation 
(QSHJE). In compensation, we incorporate from the beginning 
of the formalism hidden parameters which are 
represented by the non-classical integration constants  
appearing in the reduced action. That's precisely what we 
have done in Ref. \cite{BD1}.

Furthermore, if we would take up the Lagrangian depending on  
$(x,\dot{x},\ddot{x},\dot{\ddot{x}})$ and keep the definition 
\begin {equation}
S=\int L\ dt  \; ,
\end {equation}
the quantum equation of motion, deduced by appealing to the least 
action principle, is 
\begin {equation}
{d^3 \over dt^3}{\partial L \over \partial \dot{\ddot{x}}}- 
{d^2 \over dt^2}{\partial L \over \partial \ddot{x}}+{d \over dt}
{\partial L \over \partial \dot{x}}-{\partial L \over \partial x}=0\;.  
\end {equation}
First, we see that if the dependence on $\dot{\ddot{x}}$ 
of $L$ is not linear, we will obtain an equation of sixth order. 
This is not compatible with the QSHJE which indicates that the 
fundamental law of motion must be a forth order equation 
\cite{BD1}. In addition, the  corresponding Hamiltonian 
can be constructed as follows: we calculate the total 
derivative with respect to $t$ of $L$, and look for the 
existence of any constant of motion with the 
use of (5) in the stationary case. We get 
\begin {eqnarray}
H=\left({\partial L \over \partial \dot{x}}-{d \over dt}{\partial L 
\over \partial \ddot{x}}+{d^2 \over dt^2}{\partial L \over \partial 
\dot{\ddot{x}}} \right) \dot{x} \hskip15mm&& \nonumber\\
+ \left({\partial L \over \partial \ddot{x}}-{d \over dt}
{\partial L \over \partial \dot{\ddot{x}}}\right) \ddot{x}+
{\partial L \over \partial \dot{\ddot{x}}}\dot{\ddot{x}}-L\; ,
\end {eqnarray}
\noindent
so that 
\begin {equation}
{dH \over dt}=0\; ,
\end {equation}
if $\partial L / \partial t=0$. At this stage, many 
difficulties appear in the search of canonical equations. 
In fact, if we write the Hamiltonian as a function of 
$(x,P)$, this last set of variables will not be 
sufficient to substitute the set $(x,\dot{x},\ddot{x},
\dot{\ddot{x}})$. If we add to the set $(x,P)$ the derivatives 
$\dot{P}$ and $\ddot{P}$, we lose the symmetry between the canonical 
variables $x$ and $P$. If we write $H$ as a function of $(x,P,\dot{x},
\dot{P})$, in the classical limit $\hbar \to 0$, $P$ and $\dot{x}$ 
will form a redundant subset. It is not easy, may be impossible, to 
construct an Hamiltonian with canonical variables from which we use 
the Hamilton-Jacobi procedure to get to the third order well-known 
QSHJE. 

Now, let us consider the problem of constants which seem forming a 
redundant subset. In order to go round this problem, let us present 
the Lagrangian formulation in the following manner. We appeal to the 
quantum transformation 
$$
x \to \hat{x} \; , 
$$
introduced by Faraggi and Matone \cite{FM1,FM2}, with which the 
quantum equations take the classical forms. Then, we write the 
quantum Lagrangian in the form
\begin {equation}
\hat{L} (\hat {x}, {\dot {\hat{x}} }) = 
    {1 \over 2}m {\dot {\hat{x}} }^2 -\hat {V} (\hat {x}) \; .
\end {equation}
The hidden parameters introduced in \cite{BD1}, the energy and the 
coordinate $x$ are absorbed in $\hat{x}$. In (8), we can consider 
$\hat{x}$ and ${\dot {\hat{x}}}$ as independent variables and 
then the equation of motion,
\begin {equation}
 {d \over dt}{\partial \hat{L} \over \partial \dot{ \hat {x}}}
-{\partial \hat{L} \over \partial \hat{x}} =0 \; ,
\end {equation}
resulting from the least action principle leads to
\begin {equation}
m\ {d^2 \hat{x} \over dt^2}=-{d\hat{V} \over d\hat{x}}\; .
\end {equation}
This relation recalls us the classical Newton's law. 
As $\hat{V}(\hat{x})=V(x)$, integrating Eq. (10) gives
\begin {equation}
m\left({d \hat{x} \over dt}\right)^2+\hat{V}(\hat{x})=E \; , 
\end {equation}
\noindent 
where the integration constant $E$ is identified to the energy 
of the system because, in the classical limit $\hbar \to 0$, 
$\hat{x}$ reduces to $x$ \cite{FM1,FM2} and (11) must reproduce 
the classical law of the energy conservation. Until now, we 
have no redundant subset. Let us at present express (11) in terms of 
$x$. Again, the relation $\hat{V}(\hat{x})=V(x)$ allows us to 
write
\begin {equation}
m\ \left({dx \over dt}\right)^2 \left({\partial \hat{x} \over 
\partial x}\right)^2+V(x)=E\; . 
\end {equation}
Taking into account the expression
\begin {equation}
{\partial \hat{x} \over \partial x}=
 {\partial S_0 /\partial x \over \sqrt{2m(E-V(x))} }\; ,
\end {equation}
which we deduce from Eq. (8) in \cite{BD1} or (56) in \cite{FM2}, 
Eq. (12) leads to
\begin {equation}
\dot{x}{\partial S_0 \over \partial x}=2[E-V(x)]\; .
\end {equation}
This equation is exactly the same as the one we get in 
\cite{BD1} by expressing the Lagrangian in terms of $(x, 
{\dot x})$ and hidden parameters. 

We stress that it is also possible to reproduce Eq. (14) 
with an Hamiltonian formulation. In fact, as shown by 
Faraggi and Matone \cite{FM1,FM2}, the QSHJE can be 
written as 
\begin {equation}
E={1 \over 2m} \left( {\partial S_0 \over \partial x}\right)^2\          
\left( {\partial x \over \partial \hat{x}}\right)^2+V(x)\; .
\end {equation}
\noindent
Substituting $E$ by the Hamiltonian $H$ and 
${\partial S_0 / \partial x}$ by $P$, we get 
\begin {equation}
H={P^2 \over 2m} \left( {\partial x \over \partial 
\hat{x}}\right)^2 +V(x)\; ,
\end {equation}
which leads to the canonical equation
\begin {equation}
\dot{x}= {\partial H \over \partial P }
        ={P \over m} 
         \left( {\partial x \over \partial \hat{x}}\right)^2 \; ,
\end {equation}
since $\hat{x}$ does not depend on the derivative of $x$. By 
using this last equation in (15), we reproduce (14). As for 
the quantum version of Jacobi's theorem \cite{BD1}, relations 
(16) and (17) constitute another proof that we can reproduce 
our fundamental result, Eq. (14), without appealing to any 
Lagrangian formulation.

In the last reservation expressed by Floyd, it is stated that our 
use of the quantum coordinate implies that classical mechanics 
would be consistent with the quantum equivalence postulate (QEP). 
In his reasoning, he considered two classical systems, 
${\cal A}_{classical}$ and ${\cal B}_{classical}$, and their 
corresponding quantum systems, ${\cal A}_{quantum}$ and 
${\cal B}_{quantum}$. According to the QEP, ${\cal A}_{quantum}$ 
and ${\cal B}_{quantum}$ can be connected by coordinate 
transformation. It follows that ${\cal A}_{classical}$ and 
${\cal B}_{classical}$ can be also connected because the quantum 
transformation would relate ${\cal A}_{quantum}$ and 
${\cal A}_{classical}$ as well as ${\cal B}_{quantum}$ and 
${\cal B}_{classical}$ consistent with QEP. 
We would like to emphasize that we have not assumed that 
transformation (8) in \cite{BD1} follow from the QEP. This 
equation is just a step which allow us to reduce the QSHJE 
to the classical form in order to apply classical laws to 
the quantum motion. Of course, this step is different from 
the maps which we consider when we connect different states.

Now, let us discuss the validity of Floyd's version of 
Jacobi's theorem. We stress that in classical 
mechanics this theorem is a consequence of a particular 
canonical transformation which makes the new Hamiltonian 
vanish. The resulting Hamilton-Jacobi equation is a first 
order one. In quantum mechanics, if we use the coordinate 
$\hat{x}$ with which the quantum laws take the classical forms, 
we can reproduce the procedure of the canonical transformation 
making the new Hamiltonian vanish and get to the Jacobi's 
theorem so that 
\begin {equation}
(t-t_0)_1=\left[{\partial \hat{S_0}(\hat{x})\over \partial E} 
\right]_{\hat{x}=cte}\; .
\end {equation}
The resulting QSHJE, in which $\hat{x}$ is the variable, will be 
a first order one. With regard to the Jacobi's theorem as 
written by Floyd \cite{Floyd2}, 
\begin {equation}
(t-t_0)_2=\left[{\partial S_0(x) \over \partial E}\right]_{x=cte}\; ,
\end {equation}
we observe that there is no procedure which starts from an 
Hamiltonian formulation and leads to the third order 
well-known QSHJE. As $\hat{S_0}(\hat{x})=S_0(x)$, the difference 
between (18) and (19) can be showed in the following 
relations 
\begin {eqnarray}
(t-t_0)_2 & = & \left[{\partial \over \partial E}\  S_0(x,a,
b,E)\right]_{x=cte}  \nonumber\\
 & = & \left[{\partial \over \partial E}\  
\hat{S}_0[\hat{x}(x,a,b,E),E]\right]_{x=cte}  \nonumber\\
 & = & \left[{\partial \hat{S}_0 \over 
       \partial E} \right]_{\hat{x}=cte}
+{\partial \hat{S}_0 \over \partial 
\hat{x}} \left[{\partial \hat{x} \over 
\partial E}\right]_{x=cte}    \nonumber\\  
& = & (t-t_0)_1 + {\partial \hat{S}_0 \over \partial 
\hat{x}} \left[{\partial \hat{x} 
\over \partial E}\right]_{x=cte} \; .
\end {eqnarray}
Now, let us consider the argument proposed by Floyd to justify 
the use of its version of Jacobi's theorem. From the relation 
$S_0=S+Et$ between the reduced action $S_0(x,E,a,b)$ and 
the Hamilton's principle function $S(x,t,E,a,b)$, he 
first calculated the derivative with respect to $t$ and got 
\begin {equation}
{\partial S \over \partial t}=-E \; .   
\end {equation}  
Then, he calculated the derivative with respect to $E$ and got 
\begin {equation}
{\partial S_0 \over \partial E}={\partial S \over \partial E}
+{\partial t \over \partial E}+t \; .    
\end {equation}  
In (21 ), he considered $E$ and $t$ as independent, while in 
(22) $t$ is considered as a function of $E$. Furthermore, 
he substituted in (22) 
$\partial S / \partial E$  
by
$(\partial S / \partial t)(\partial t / \partial E)$. 
Firstly, in our point of view, when we consider $S(x,t,E,a,b)$, 
all the elements of the set $(x,t,E,a,b)$ must be seen as 
independent. Secondly, even if we suppose that $t=t(E)$, 
we can not substitute 
$\partial S / \partial E$  
by
$(\partial S / \partial t)(\partial t / \partial E)$ 
because in this case we have $S=S(x,t(E),E,a,b)$ and we must write 
\begin {equation}
{\partial S \over \partial E}=
    \left.{\partial S \over \partial E}\right|_{t=cte}
    +\left.{\partial S \over \partial t}\right|_{E=cte} 
{\partial t \over \partial E}  \; .  
\end {equation}  
Thirdly, he got ${\partial S_0 / \partial E}=t$
representing the Jacobi's theorem. We observe in 
describing the motion for any initial condition that there
is an integration constant missing from this equation. 

To conclude this discussion about Floyd's version of Jacobi's 
theorem, let us show that the trajectories depend on the 
choice of the couple of Schr\"odinger's solutions. The 
reduced action is \cite{BD1}
\begin {equation}
S_0=\hbar \arctan\left(a{\phi_1 \over \phi_2}+b\right)
+\hbar \lambda \; ,
\end {equation}
$(\phi_1,\phi_2)$ being a real set of independent solutions 
of Schr\"odinger's equation. As an 
example, we consider a free particle of energy $E$ and we 
set $k=\sqrt{2mE}/\hbar$. If we choose 
\begin {equation}
\phi_1=\sin (kx) \; , \ \ \ \ \ \ \ \phi_2=\cos (kx) \; ,
\end {equation}
and use Jacobi's theorem as proposed by Floyd, we get
\begin {equation}
t=t_0+{ma \over \hbar k}{x \over (1+b^2)\cos^2 (kx)+a^2\sin^2 (kx) 
+2ab \sin (kx) \; \cos (kx)}  \; . 
\end {equation}  
Another possible choice is 
\begin {equation}
\theta_1=\sin (kx) +g(k) \cos (kx)  \; , 
\end {equation}  
\begin {equation}
\theta_2=\cos (kx) +f(k) \sin (kx)   \; ,
\end {equation}  
where $f$ and $g$ are two arbitrary real functions of $k$ 
satisfying the condition $fg \ne 1$. We indicate that Floyd 
\cite{Floyd3} has also used linear combinations of 
Schr\"odinger's solutions with coefficients 
depending on $k$. For simplicity, we choose in what follows 
$g(k)=0$. Now, let us look for the existence of three real 
parameters $\tilde{a}$, $\tilde{b}$ and $\tilde{t}_0$ with  
which the reduced action takes the form 
\begin {equation}
\tilde{S}_0=\hbar \arctan \left( \tilde{a} 
{\theta_1 \over \theta_2}+\tilde{b}\right) + \hbar \tilde{\lambda}
\end {equation}  
as in (24), and the resulting Floyd's trajectory,  
\begin {eqnarray}
t=\tilde{t}_0+{m \tilde{a} \over \hbar k} 
\left[x-{df \over dk}\sin^2 (kx) \right] 
\left[(1+\tilde{b}^2)\cos^2 (kx) \right. \hskip35mm&& \nonumber\\ 
+(\tilde{a}^2+\tilde{b}^{2}f^{2}+f^2+2\tilde{a}\tilde{b}f)
\sin^2 (kx) \hskip20mm&& \nonumber\\
\left. +2(f+\tilde{a}\tilde{b}+\tilde{b}^{2}f) 
\sin (kx) \; \cos (kx)  \right]^{-1}  \; ,
\end {eqnarray}  
reproduces the same quantum equation as (26) for every $f(k)$. 
This implies that the right hand sides of (26) and (30) must 
be identical. For $x=0$, this identity gives $t_0=\tilde{t}_0$, 
and therefore, for $x=\pi /2k$ and $x=3\pi /2k$, we deduce  
that
\begin {equation}
a\tilde{a}\left({\pi \over 2k} -{df \over dk}\right)
  =  {\pi \over 2k}\left(\tilde{a}^2+\tilde{b}^{2}f^{2}+f^{2} 
     + 2 \tilde{a}\tilde{b}f\right)
\end {equation}  
and 
\begin {equation}
a\tilde{a}\left({3\pi \over 2k} -{df \over dk}\right)
    =   {3\pi \over 2k}\left(\tilde{a}^2+\tilde{b}^{2}f^{2}+f^{2} 
        + 2 \tilde{a}\tilde{b}f\right)
\end {equation}  
respectively. These two last equations can not be simultaneously 
satisfied unless one has $df/dk=0$. Since the function $f(k)$ is 
arbitrary, the identification of Eqs. (31) and (32) leads to 
a contradiction. Thus, we get to the unsatisfactory conclusion 
for Floyd's trajectories: the mathematical choices affect the 
physics results. This is not the case for our formulation for 
which we clearly showed \cite{BD2} for any potential that the 
trajectories are independent on the choice of the couple 
$(\phi_1,\phi_2)$.  

The ambiguity appearing in the definition of the derivative 
$\partial S_0 /\partial E$ when we consider the dependence on 
$E$ of the integration constants is pointed out by 
Faraggi-Matone \cite{FM1}. In order to allow ``a non ambiguous 
definition of time parametrization'', they suggested that all 
the terms depending on $E$ and which can be absorbed in a 
redefinition of the integration constants should not be considered 
in evaluating $\partial S_0 /\partial E$. With this proposal, we 
can indeed show that Floyd's trajectories are independent on 
the choice of the couple $(\phi_1,\phi_2)$. In fact, let us consider 
the transformation
\begin {equation}
\phi_1 \to \theta_1 = \mu \phi_1 + \nu \phi_2 \; , 
\end {equation}
\begin {equation}
\phi_2 \to \theta_2 = \alpha \phi_1 + \beta \phi_2 \; .
\end {equation}
where the real parameters $(\mu, \nu, \alpha, \beta)$ are depending 
on $E$ and satisfying the condition $\mu \beta \not= \nu \alpha$.  
If we write for any potential the new reduced action as in (29), 
with the same procedure developed in Ref. \cite{BD2}, we can find 
$\tilde{a}=\tilde{a}(a, b,\mu, \nu, \alpha, \beta)$ 
and 
$\tilde{b}=\tilde{b}(a, b, \mu, \nu, \alpha, \beta)$
in such a way as to guarantee that 
$\partial S_0 /\partial x = \partial \tilde{S}_0 /\partial x$.
This equality implies that, up to an additive constant, $S_0$ and 
$\tilde{S}_0$ are identical. In other words, we can write
\begin {equation}
\tilde{S}_0=\hbar \arctan\left[a(\tilde{a},\tilde{b},\mu, \nu, \alpha, \beta)  
          {\phi_1 \over \phi_2} +
          b(\tilde{a},\tilde{b},\mu, \nu, \alpha, \beta)\right] \; ,
\end {equation}
where we have omitted the additive constant. According to Faraggi-Matone's 
proposal, it follows that
\begin{equation}
{\partial S_0 \over \partial E }= 
           {\partial \tilde{S}_0 \over \partial E} \; .
\end{equation}
However, this procedure of evaluating $\partial S_0 / \partial E$ leads 
to some unsatisfactory results. As an example, if we calculate the time 
reflection for a semi-infinite rectangular barrier \cite{Floyd3}, we get 
a vanishing value. Another unsatisfactory aspect of this procedure 
appears when we consider the free case with $E=0$ as a limit 
from arbitrary $E$. In fact, if we want to reproduce the two independent 
solutions $\phi_1^0 = x$ and $\phi_2^0 = 1$ of the free case 
\cite{FM1,FM3} from the solutions (25) when we consider the limit 
$E \to 0$, we must rescale $\phi_1(x)$ as follows
\begin{equation}
\phi_1 = k^{-1} \; \sin(kx) \; .
\end{equation}
%
If we want to keep this possibility of reproducing the free case 
in the limit $E \to 0$ when we apply Jacobi's theorem, we must 
write explicitly the factor $k^{-1}$ in the expression of the reduced 
action and this will give rise to a further term in the right hand side 
of (26). It is clear that this creates confusion in the definition 
of time parametrization. 

\bigskip
\bigskip

We would like to thank E.R. Floyd for interesting discussions and 
encouragements despite our disagreements about the formulation 
of trajectory representation of quantum mechanics.

  
\vskip\baselineskip
\noindent
{\bf REFERENCES}

\end {document}